\documentclass{aa}  

\usepackage{graphicx}
\usepackage{txfonts}

\usepackage[hidelinks]{hyperref}
\let\linenumbers\relax

\begin{document}

\title{Detection and characterisation of Starlink VHF TT\&C emissions at 137.055~MHz with the 21 CentiMeter Array}

%
%
%

   \author{Yanbin Yang\inst{1,2,3}      
        \and Feiyu Zhao\inst{1,2}
        \and Quan Guo\inst{4}\corrauth{guoquan@shao.ac.cn}
        \and Junhua Gu\inst{5}\corrauth{jhgu@nao.cas.cn}
        \and Qian Zheng\inst{4}\corrauth{qzheng@shao.ac.cn}
        \and Yan Huang\inst{5} 
        \and Yun Yu\inst{1,4}
        \and Ming Jiang\inst{6}
        \and Jianfeng Wang\inst{7,2}
        \and Xiaoming Zhang\inst{7,2}}

   \institute{Shanghai Astronomical Observatory, Chinese Academy of Sciences, 80 Nandan Road, Shanghai 200030, China 
   \and University of Chinese Academy of Sciences, No.1 Yanqihu East Road, Beijing 101408, China
   \and School of Physical Science and Technology, ShanghaiTech University, Shanghai 201210, China
   \and State Key Laboratory of Radio Astronomy and Technology,  Shanghai Astronomical Observatory, Chinese Academy of Sciences, 80 Nandan Road, Shanghai 200030, China
   \and State Key Laboratory of Radio Astronomy and Technology, National Astronomical Observatories, Chinese Academy of Sciences, 20A Datun Road, Beijing 100101, China
   \and National Key Laboratory of Radar Signal Processing, Xidian University, Xi'an, 710071, China
   \and National Astronomical Observatories, Chinese Academy of Sciences, 20A Datun Road, Beijing 100101, China}

   \date{Received September 30, 20XX}

 
  \abstract
   {Satellite emissions at low frequencies can interfere with astronomical observations. Observations with the Engineering Development Array version 2 (EDA2) have revealed recurrent Starlink emissions near 137~MHz, attributed to very high frequency (VHF) telemetry, tracking, and command (TT\&C) transmissions.}
   {We aim to characterise Starlink VHF emissions detected with the 21 CentiMeter Array (21CMA) to better assess their potential impact on future low-frequency radio astronomical observations and inform the development of methods to mitigate the resulting interference.}
   {We searched five coordinated VHF channels in observations obtained on 15 April 2025 (UTC), demodulating and decoding two LoRa packets. A payload-derived state vector was compared with public orbital catalogues to identify the transmitter. Differential Doppler measurements tested the association between the two events. We examined orbital visibility and evolution to investigate the limited number of detections, reanalysed published EDA2 detections across orbital phases, and estimated the stronger event's flux density.}
   {Two valid packets are detected at 137.055~MHz, each lasting 1.027584~s and separated by 99.987~s. A blind comparison with 11\,305 active satellites identifies STARLINK-32838 (NORAD ID 62827) as the transmitter, with differential Doppler measurements linking both events to the same spacecraft. Orbital analysis shows that the satellite was undergoing orbit raising and that the limited number of detections was partly attributable to its being below the horizon for most of the observing time. Re-analysis of EDA2 detections indicates that such emissions are not confined to orbit raising. The stronger event's estimated flux density is 9.23~MJy at 1000~km, consistent with previous measurements at the order-of-magnitude level.}
   {Signal demodulation and decoding, combined with orbital analysis, can improve the identification and characterisation of satellite-related radio frequency interference. The detection of similar VHF emissions across different orbital phases indicates that their potential interference with low-frequency radio astronomy should not be considered solely in the context of orbit raising.}


\titlerunning{Starlink VHF TT\&C emissions at 137.055~MHz}
   \maketitle

\section{Introduction}

Radio frequency interference (RFI) is a persistent challenge for low-frequency radio astronomy, particularly for highly sensitive instruments such as the Square Kilometre Array Low (SKA-Low), which aims to detect extremely faint redshifted 21-cm emission from the epoch of reionisation and the cosmic dawn \citep{SKA_low}. Various approaches, including the establishment of radio-quiet zones, the use of natural terrain shielding, and interference detection and flagging, have been adopted to mitigate RFI. However, while remote observatory locations and topographical shielding can reduce terrestrial interference, they offer less protection against signals from aircraft and satellites. With the continued growth of low-Earth-orbit (LEO) satellite constellations, the threat from the sky warrants increasing attention.

Recent observations have demonstrated that Starlink satellites emit signals at frequencies far below their commercial broadband communication bands. Using the Low-Frequency Array (LOFAR), \citet{DiVruno} detected unintended electromagnetic radiation (UEMR) between 110 and 188~MHz, while \citet{Bassa} subsequently reported stronger broadband UEMR from second-generation Starlink satellites. The Engineering Development Array version 2 (EDA2), an SKA-Low prototype station in Western Australia, has also recorded widespread Starlink emissions at low frequencies \citep{EDA2_2023, EDA2_2025}. A related search for Starlink UEMR has been conducted using the 21 CentiMeter Array (21CMA) in China \citep{21CMA_UEMR}.

These findings have motivated efforts to better characterise and mitigate satellite interference. In addition to conventional RFI excision, coordination between observatories and satellite operators offers a potential means of reducing certain transmissions during sensitive observations. For example, \citet{Nhan} demonstrated that Starlink commercial downlink channels could be temporarily redirected or disabled during close passages to the Green Bank Telescope. The SKAO has also reported that it is testing and implementing Boresight Avoidance and has already trialled this approach with SpaceX using the Onsala Twin Telescopes in Sweden \footnote{\url{https://www.skao.int/en/science-users/642/science-user-qa}}. However, these measures do not directly address UEMR from onboard satellite electronics.

Beyond UEMR, Starlink satellites also produce intended electromagnetic radiation (IEMR) within the low-frequency observing range. Using the EDA2, \citet{EDA2_2023} reported strong, approximately 100~s periodic emissions near 137~MHz, which were associated with Starlink very high frequency (VHF) telemetry, tracking, and command (TT\&C) transmissions. Subsequent EDA2 observations constrained similar narrowband emissions to approximately 137.05~MHz \citep{EDA2_2025}. The Federal Communications Commission (FCC) has authorised Starlink VHF TT\&C downlinks on several channels within 137--138~MHz, including 137.055~MHz \citep{fcc2024starlinkbeacon}. Recent ground-station measurements have further demonstrated the reception of LoRa-modulated Starlink VHF packets \citep{IEEEStarlinkVHF}. While these studies have established the presence of Starlink VHF emissions, a more comprehensive characterisation of their signal properties, transmitting satellites, orbital context, and received signal strengths would help improve our understanding of these emissions and their implications for low-frequency radio astronomy.
 
In this work, two Starlink VHF TT\&C events are detected at 137.055~MHz in raw voltage observations obtained with the 21CMA on 15 April 2025 (UTC). We characterise their LoRa waveform structure and recover complete data packets through physical-layer (PHY) demodulation and decoding. By interpreting information contained in the decoded payloads and comparing it with publicly available satellite orbital data, we identify the transmitting spacecraft as Starlink-32838 (NORAD ID 62827) and examine the consistency of the two detections. We further investigate their observing geometry and orbital context, estimate the received flux density, and re-examine previously reported EDA2 detections to explore the occurrence of VHF emissions across different orbital phases.

The structure of this paper is as follows. Section~\ref{section2} describes the observations, systematic search, and demodulation of the two detected events. Section~\ref{section3} identifies the transmitting satellite through orbital comparisons and examines the association between the two events using differential Doppler measurements. Section~\ref{section4} investigates orbital visibility and evolution, including a re-analysis of previous EDA2 detections across different orbital phases. Section~\ref{section5} presents the flux-density estimation. Discussion and conclusions are in Section~\ref{section6} and Section~\ref{section7}, respectively.

\section{Observations and data analysis}
\label{section2}

The 21CMA is a low-frequency radio interferometer located in Ulastai, Xinjiang, China. It consists of 81 pods arranged along two perpendicular arms oriented east-west and north-south. Each pod comprises 127 single-polarisation log-periodic antennas fixed towards the North Celestial Pole (NCP).

On 15 April 2025 (UTC), we used the E08 pod ($42.92424^\circ$ N, $86.72398^\circ$ E;  WGS84 ellipsoidal height 2607~m) to conduct observations originally aimed at searching for UEMR from Starlink satellites at low frequencies. The raw voltage data were collected over five observing windows (see Table~\ref{table:windows}) and cover a filtered frequency range of 50--200~MHz. Although no convincing Starlink UEMR was detected, ORBCOMM transmissions were successfully decoded, demonstrating the suitability of the data for narrowband signal analysis \citep{21CMA_UEMR}.

\begin{table}[ht!]
\caption{Observing windows on 15 April 2025 (UTC).}                 
\label{table:windows}   
\centering
\begin{tabular}{cccc}     
\hline\hline          
Window & Start Time & End Time & Duration (s) \\         
\hline                     
W1 & 11:39:52 & 11:51:06 & 674 \\
W2 & 12:59:53 & 13:09:41 & 588 \\
W3 & 13:59:56 & 14:14:34 & 878 \\
W4 & 16:35:53 & 16:38:44 & 171 \\
W5 & 23:34:52 & 23:38:00 & 188 \\
\hline
Total & -- & -- & 2499 \\
\hline                               
\end{tabular}
\end{table}

During a subsequent characterisation of the low-frequency radio environment at the 21CMA site using the same data, we identified an initially unknown narrowband event near 137.055~MHz. Its high-resolution spectrogram revealed a sequence of chirps characteristic of LoRa modulation. We refer to this signal as Event~1. PHY demodulation and analysis of the recovered packet indicated that the event was consistent with a Starlink VHF TT\&C downlink. A systematic search of the complete data identified one additional event with similar waveform and packet properties, denoted Event~2.

\subsection{LoRa waveform characteristics}

Starlink second-generation satellites are authorised to operate backup VHF TT\&C beacons on five coordinated space-to-Earth channels within the 137--138~MHz band, including 137.055~MHz. When active, these beacons are designed to transmit a telemetry and tracking message once every 100~s on average \citep{fcc2024starlinkbeacon}. Independent ground-based measurements have identified these transmissions as using a LoRa-based waveform derived from the Swarm SpaceBEE system \citep{IEEEStarlinkVHF,kenny2025starlink}. This motivates a comparison of our detected events with the characteristic time--frequency structure of LoRa signals.

The LoRa PHY uses chirp spread-spectrum modulation, in which the signal frequency sweeps linearly with time. Information is encoded by cyclically shifting a reference chirp. In a dynamic spectrum, an upchirp appears as a track whose frequency increases with time, whereas a downchirp sweeps in the opposite direction. The packet structure relevant to this work consists of a repeated-upchirp preamble, two synchronisation symbols, a start-frame delimiter (SFD) containing downchirps, and the subsequent PHY data-symbol region. In both events, the PHY data-symbol region contains an explicit PHY header, a payload, and a payload cyclic redundancy check (CRC).

To characterise the time--frequency morphology of the detected signals and compare the two events at different temporal scales, we constructed multi-scale dynamic spectra from the downconverted complex baseband data. To remove the frequency-dependent spectral baseline, we selected two 1~s off-event segments for each packet, one before and one after, each separated from the packet by a 0.30~s guard interval. For each frequency bin, we calculated the median off-event power spectral density in decibels (dB) and subtracted it from the corresponding dynamic spectrum. At each scale, both events were processed using the same parameters and displayed with a common colour scale, allowing direct comparison of their relative spectral contrasts.

\begin{figure*}
        \centering
        \includegraphics[width=\hsize]{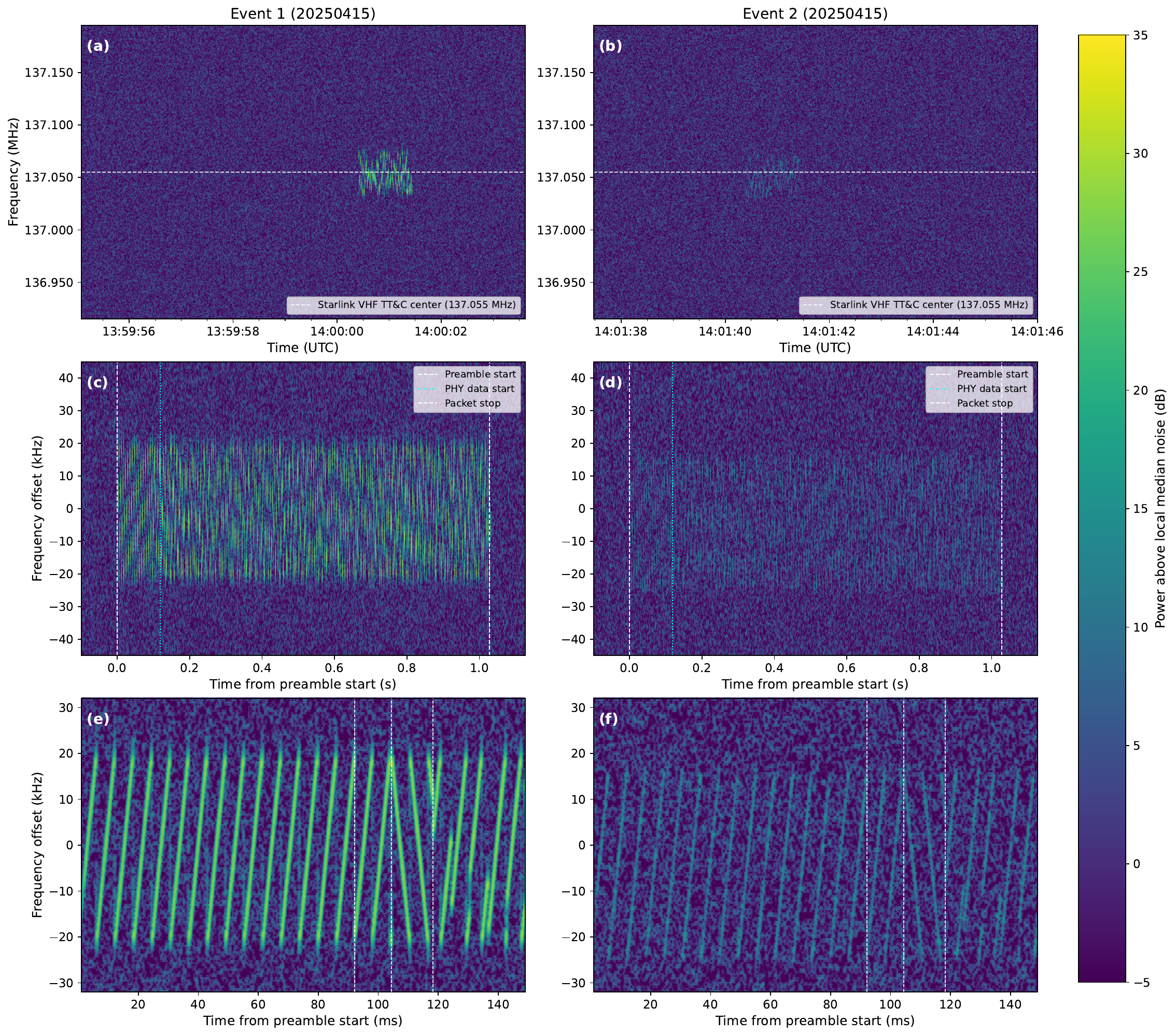}

        \caption{Dynamic spectra of Event~1 (left) and Event~2 (right) at three progressively narrower time--frequency ranges. Top: event locations in UTC and absolute frequency; the horizontal white dashed line marks the nominal 137.055~MHz Starlink VHF TT\&C channel centre. Middle: complete packets with 0.10~s of surrounding background on either side, referenced to the preamble onset in time and to 137.055~MHz in frequency. The white dashed lines mark the packet boundaries, and the cyan dotted line marks the beginning of the post-SFD PHY data-symbol region. Bottom: the first 150~ms of each packet, with dashed boundaries separating the preamble, synchronisation symbols, SFD, and PHY data-symbol region. Power is expressed in dB relative to the median off-event background in each frequency bin, using a common colour scale throughout. All spectra use 1024-sample Hann windows at 480~ksample~s$^{-1}$, giving a window duration of 2.133~ms and a frequency bin spacing of 468.75~Hz.}
        \label{pic:waterfall}
    \end{figure*}

Figure~\ref{pic:waterfall} presents the two events at three progressively narrower time--frequency ranges. The upper panels locate the events in UTC and frequency, with Event~2 exhibiting a substantially lower spectral contrast than Event~1. The middle panels display the complete packets, with the preamble start defined as $t=0$ and the frequency offset expressed relative to the nominal 137.055~MHz channel centre. Both events extend for more than 1~s and exhibit similar occupied bandwidths. The lower panels show the first 150~ms in greater detail, revealing the characteristic LoRa packet structure. Each packet begins with 15 repeated preamble upchirps, followed by two shifted synchronisation symbols and an SFD containing 2.25 downchirps. After the SFD, chirps with varying cyclic shifts mark the beginning of the PHY data-symbol region. Despite the lower spectral contrast of Event 2, the same packet structure is evident in both events.

\subsection{Demodulation}

Although both events exhibit time--frequency morphologies consistent with the reported LoRa-based Starlink VHF waveform, morphology alone is insufficient to establish that they are valid packets or to distinguish them conclusively from visually similar RFI. Demodulation enables the complete packets to be recovered and their PHY validity and payload fields to be examined, providing stronger evidence for identifying them as Starlink VHF packets. 

In this work, we developed a dedicated Starlink VHF TT\&C demodulation pipeline as part of \texttt{LFdemod}, an open-source Python toolkit for the offline demodulation and diagnostic analysis of low-frequency radio signals \citep{lfdemod}. The pipeline uses \texttt{lora-phy} as its underlying LoRa PHY decoding backend \citep{loraphy}. The real-valued voltage data were digitally downconverted and channel-filtered around the nominal channel centre of 137.055~MHz. The resulting complex baseband data were resampled to 83.333~ksample~s$^{-1}$, corresponding to twice the LoRa bandwidth. The LoRa PHY demodulator was configured with a bandwidth (BW) of 41.667~kHz, spreading factor (SF) 8, a 15-symbol preamble, sync word \texttt{0x12}, an explicit PHY header, and low-data-rate optimisation enabled. The nominal symbol duration was $T_{\mathrm{sym}}=2^{SF}/BW=6.144$~ms. 

The PHY data start and carrier-frequency offset were estimated from the repeated preamble chirps and the SFD. The preamble start was then inferred by subtracting $(15+2+2.25)T_{\mathrm{sym}}$ from the PHY data start, accounting for the preamble upchirps, two sync symbols, and the SFD. Following dechirping and symbol estimation using a fast Fourier
transform (FFT), the recovered symbols were Gray remapped, diagonally deinterleaved, Hamming decoded, and dewhitened. The explicit PHY header was parsed and its checksum verified, after which the payload was reconstructed and its CRC was checked. The decoded header and adopted PHY configuration yielded 148 PHY data symbols, giving a calculated packet duration of $(15+2+2.25+148)T_{\mathrm{sym}}=1.027584$~s.

Despite the lower spectral contrast of Event 2, both events were decoded into complete packets with valid PHY headers and payload CRCs. Each packet contains 148 PHY data symbols and an 81-byte payload encoded at a coding rate of 4/5. The byte positions and field names used to parse the 81-byte payloads were adopted from the independently reverse-engineered format reported by \citet{kenny2025starlink}. All multi-byte quantities were interpreted in little-endian order. The complete decoded payloads, represented as hexadecimal byte strings, are available here \footnote{\url{https://github.com/JinYi1108/21CMA_Starlink_VHF/tree/main/paper_results/demodulation}}.

The two payloads have the same internal protocol identifier, 5927, and packet type value, \texttt{0xCC}. Their message numbers increase consecutively from 3441186 to 3441187, while the decoded UTC timecode increases from 1744725561 to 1744725661, corresponding to an interval of exactly 100~s. In the recorded voltage data, the two observed preamble start times are separated by 99.987~s, only 0.013~s shorter than the 100~s interval. These agreements strongly support the interpretation that the two events are consecutive transmissions from the same source. Selected quantities relevant to packet validation and comparison between the two events are listed in
Table~\ref{table:event_properties}.

\begin{table*}[ht!]
\caption{Observed and decoded properties for the two Starlink VHF events.}
\label{table:event_properties}
\centering
\begin{tabular}{lcc}
\hline\hline
Parameter & Event 1 & Event 2 \\
\hline

\multicolumn{3}{l}{\textit{Observed properties}} \\

Preamble start time (UTC) &
14:00:00.413 &
14:01:40.400 \\

Time relative to Event 1 (s) &
0 &
99.987 \\

Calculated packet duration (s) &
1.027584 &
1.027584 \\

Nominal channel centre (MHz) &
137.055 &
137.055 \\

Measured centre frequency (MHz) &
137.053910 &
137.050410 \\

Carrier-frequency offset (kHz) &
$-1.090$ &
$-4.590$ \\

\hline
\multicolumn{3}{l}{\textit{PHY decoding and validation}} \\

Recovered sync word &
\texttt{0x12} &
\texttt{0x12} \\

Calculated PHY data symbols &
148 &
148 \\

PHY-header checksum &
Valid &
Valid \\

Coding rate &
4/5 &
4/5 \\

Payload length (bytes) &
81 &
81 \\

Received payload CRC &
\texttt{0xA39C} &
\texttt{0x4FC8} \\

Payload CRC status &
Valid &
Valid \\

\hline
\multicolumn{3}{l}{\textit{Parsed payload fields}} \\

Message number &
3441186 &
3441187 \\

Internal protocol identifier &
5927 &
5927 \\

Packet type &
\texttt{0xCC} &
\texttt{0xCC} \\

UTC timecode &
1744725561 &
1744725661 \\

\hline
\end{tabular}

\tablefoot{
Only selected quantities relevant to packet validation and event comparison are shown. The preamble start times combine the UTC timing of the corresponding recordings with the file-relative preamble onsets determined by the demodulator. The reported 81-byte length refers only to the LoRa payload. The internal protocol identifier here is labelled 'Spacecraft ID' in the independently reverse-engineered packet description. It is an internal identifier and is not a NORAD catalogue number.}
\end{table*}

\subsection{Systematic search for Starlink VHF events}

To determine whether additional signals of the same type were present in the data, we systematically searched all five observing windows listed in Table~\ref{table:windows}. The search covered the five coordinated Starlink space-to-Earth VHF downlink channels centred at 137.055, 137.145, 137.8463, 137.8763, and 137.9663~MHz \citep{fcc2024starlinkbeacon}.

Candidate events were first identified using sparsely sampled, 65\,536-point Hann-windowed FFTs of the raw voltage data, calculated every 25~ms. For each target channel, the mean power of the three strongest bins within $\pm35$~kHz of its nominal centre was compared with the median power over 136.98--138.04~MHz, excluding $\pm50$~kHz around all five target channels. The resulting target-to-background power ratio was expressed in dB and then normalised, separately for each target channel, by subtracting its temporal median over all sampled times in that file, thereby suppressing stationary spectral structure. Detections separated by no more than 0.6~s were grouped within each channel, with each group limited to 1.2~s. Candidate events were retained if they contained at least one detection of 8~dB or greater, three detections of at least 4~dB within 1.2~s, or two detections of at least 6~dB within 0.8~s.

A separate time mask for possible receiver overload was constructed from raw analogue-to-digital (ADC) converter amplitude statistics and broadband spectral enhancements over 75--200~MHz. Snapshots were flagged based on ADC-rail or near-full-scale sample fractions, jointly elevated high-percentile amplitudes and root-mean-square (RMS) levels, or simultaneous enhancements in integrated broadband power and multiple 1~MHz sub-bands. Flagged snapshots separated by no more than 0.2~s were merged into intervals, which were then extended by 0.1~s on each side; overlapping candidates were excluded because their narrowband origin could not be assessed reliably.

To reduce unnecessary digital downconversion (DDC) and full decoding, the remaining candidates were examined for LoRa-like chirp morphology directly in the raw voltage data. For each candidate, one to four 20~ms segments centred on distinct energy-detection times, including the first and strongest detections, were examined. Local short-time Fourier transforms were calculated using a 131\,072-point Hann window and a 65\,536-sample hop, retaining a 120~kHz band centred on the target channel. After median subtraction along both the time and frequency axes, the residual spectra were converted to robust $z$-scores using the median absolute deviation. Within each time--frequency region, we searched for 3~ms positive-slope tracks within 30\% of the expected LoRa upchirp slope of $6.78~\mathrm{kHz\,ms^{-1}}$. The strongest horizontal and negative-slope tracks in the same region were also measured as controls for stationary narrowband features and opposite-direction frequency sweeps. A candidate passed if its strongest positive-slope track had a mean score of at least 1.5, at least 40\% of the track samples had $z\ge2$, and its score exceeded the larger of the two control scores by at least 0.6.

To account for potentially long packets and uncertainties in the coarsely estimated candidate times while limiting computational cost, passing candidates were subjected to a DDC-based synchronisation test using segments of at least 1.6~s, which searched for the 15-upchirp preamble, SFD, and sync word \texttt{0x12}. Matched candidates were then fully decoded using conservative 5.5~s windows. Final validation required a valid explicit PHY header and payload CRC, together with a match to the adopted Starlink VHF packet profile. Detections were merged as the same physical event when they occurred within the same observing window, had recovered preamble times within 25~ms, and had either identical decoded payloads or measured centre frequencies within 5~kHz.

Across all observations, the initial energy search yielded 1\,450 candidates, of which 257 overlapped the overload mask. Of the remaining 1\,193 candidates, only two passed the chirp-morphology screening and subsequent synchronisation and decoding stages. These were Event~1 and Event~2. 

\section{Identification of the transmitting satellite}
\label{section3}

The demodulation results presented in Section~\ref{section2} establish both events as valid Starlink VHF TT\&C packets. However, the internal protocol identifier contained in the payload cannot be mapped directly to a NORAD catalogue number and therefore does not identify the transmitting satellite. In this section, we further interpret the payload state information, blindly compare the six-dimensional state recovered from Event~1 with the complete catalogue of active satellites at the time of observation, and use the differential Doppler shift between the two events as a consistency test. These analyses identify the transmitter of Event~1 as STARLINK-32838 (NORAD ID 62827) and associate Event~2 with the same satellite.

\subsection{Payload timing and state-vector interpretation}

To identify the transmitting satellite, we examined the timing and candidate state-vector fields in the decoded payloads. According to \citet{kenny2025starlink}, the payload format contains two distinct time representations, neither of which should be confused with the observational preamble UTC times listed in Table~\ref{table:event_properties}. Bytes 15--18 were interpreted as a 32-bit UTC timecode expressed as seconds since the Unix epoch. They correspond to 2025-04-15 13:59:21 UTC for Event~1 and 2025-04-15 14:01:01 UTC for Event~2, giving an interval of exactly 100~s. Although this progression is consistent with the nominal transmission cadence, the specific operational meaning of this timecode is unknown; in particular, we do not assume that it represents the emission time, reception time, or epoch of the satellite state vector.

A second time representation is contained in bytes 33--38. Following the reverse-engineered format, bytes 33--34 were interpreted as the GPS week number and bytes 35--38 as the GPS time of week in units of 0.01~s. After accounting for the 18~s GPS--UTC offset, the Event~1 fields give a GPS-derived state epoch of 2025-04-15 14:00:00.250 UTC. We adopt this as the nominal epoch associated with the position and velocity fields below.

The physical meaning of bytes 39--62 remained unidentified in the independent reverse-engineering report. In this work, we hypothesised that these 24 bytes encode a six-dimensional satellite state vector. They were divided into six consecutive signed 32-bit integers and scaled by $10^{-5}$. The first three quantities were interpreted as the Earth-centred, Earth-fixed (ECEF) position $(X,Y,Z)$ in kilometres, and the remaining three as the corresponding ECEF velocity $(V_X,V_Y,V_Z)$ in $\mathrm{km\,s^{-1}}$. For Event~1, this interpretation gives a position magnitude of 6689.939~km and a velocity magnitude of 7.427~$\mathrm{km\,s^{-1}}$, both physically plausible for a LEO satellite. The decoded timing and state quantities are listed in Table~\ref{table:payload_state}. Their physical plausibility alone does not identify the spacecraft, and the state-vector hypothesis is therefore evaluated through a blind comparison with the complete active-satellite catalogue in Section~\ref{subsect:blind_comparison}.

\begin{table*}[ht!]
\caption{Payload-derived timing and state parameters for the two
Starlink VHF events.}
\label{table:payload_state}
\centering
\begin{tabular}{lccc}
\hline\hline
Parameter & Payload bytes & Event 1 & Event 2 \\
\hline

\multicolumn{4}{l}{\textit{Timing fields}} \\

UTC timecode &
15--18 &
1744725561 &
1744725661 \\

UTC timecode converted to UTC &
-- &
2025-04-15 13:59:21 &
2025-04-15 14:01:01 \\

GPS week &
33--34 &
2362 &
0 \\

GPS time of week (s) &
35--38 &
223218.25 &
19.00 \\

GPS-derived state epoch (UTC) &
-- &
2025-04-15 14:00:00.250 &
Undefined \\

\hline
\multicolumn{4}{l}{\textit{State-vector fields}} \\

$X$ (km) &
39--42 &
345.38784 &
0 \\

$Y$ (km) &
43--46 &
4474.86109 &
0 \\

$Z$ (km) &
47--50 &
4961.00917 &
0 \\

$V_X$ ($\mathrm{km\,s^{-1}}$) &
51--54 &
$-6.36108$ &
0 \\

$V_Y$ ($\mathrm{km\,s^{-1}}$) &
55--58 &
3.06258 &
0 \\

$V_Z$ ($\mathrm{km\,s^{-1}}$) &
59--62 &
$-2.30717$ &
0 \\

Position magnitude (km) &
-- &
6689.93920 &
0 \\

Velocity magnitude ($\mathrm{km\,s^{-1}}$) &
-- &
7.42737 &
0 \\

\hline
\end{tabular}

\tablefoot{
Byte positions are one-based. All multi-byte quantities were decoded in little-endian order. The observational preamble UTC times are listed in Table~\ref{table:event_properties}.}
\end{table*}

For Event~2, the GPS week field is zero, while the GPS time-of-week field contains only 19~s; together they do not define a physically meaningful epoch for the observation date. In addition, all 24 bytes assigned to the position and velocity state vector are zero. The Event~2 packet nevertheless has a valid explicit PHY header and payload CRC, indicating that these values are present in the recovered payload rather than being an evident consequence of a failed decode. The reason why the Event~2 packet does not contain a usable time-tagged state vector is unknown. Consequently, only Event~1 can be used for an independent satellite search, while the association of Event~2 with the same satellite is examined separately in Section~\ref{subsect:doppler}.

\subsection{Blind matching against the active satellite catalogue}
\label{subsect:blind_comparison}

The Event~1 state vector was compared with every object in a snapshot of the complete CelesTrak \footnote{\url{https://celestrak.org/}} active satellite catalogue obtained on 15 April 2025, without assuming any prior association with Starlink satellites. The snapshot contained the then-latest two-line element (TLE) provided by CelesTrak for each of 11\,305 unique NORAD objects.

For each catalogue object, its TLE was propagated with SGP4 using \texttt{Skyfield} \citep{skyfield}, and the resulting position and velocity were expressed in the International Terrestrial Reference System (ITRS), an ECEF frame. Taking the nominal epoch (GPS-derived) as $t_0$, we allowed an independent time offset $\Delta t$ for each satellite within $-180~{\mathrm{s}}\leq\Delta t\leq 180~{\mathrm{s}}$.

The time offset was refined using the bounded Brent minimisation method. At each trial time, the position and velocity residuals were calculated as
\begin{equation}
    \Delta r(\Delta t)=\left\|\boldsymbol r_{\mathrm{TLE}}(t_0+\Delta t) - \boldsymbol r_{\mathrm{payload}}\right\|,
\end{equation}
and
\begin{equation}
 \Delta v(\Delta t)=   \left\|
\boldsymbol v_{\mathrm{TLE}}(t_0+\Delta t)
-\boldsymbol v_{\mathrm{payload}}
\right\|.
\end{equation}
Here, $\boldsymbol r_{\mathrm{TLE}}$ and $\boldsymbol v_{\mathrm{TLE}}$ denote the SGP4-propagated position and velocity, while $\boldsymbol r_{\mathrm{payload}}$ and $\boldsymbol v_{\mathrm{payload}}$ are the corresponding vectors recovered from Event~1; the residuals are Euclidean norms in the ITRS frame. The time adjustment was introduced to accommodate along-track orbital phase errors that may accumulate when propagating a TLE from its epoch to the nominal epoch.

The position and velocity residuals were combined to define a joint matching score:
\begin{equation}
    J(\Delta t)=
\Delta r(\Delta t)^2+
\left[W\Delta v(\Delta t)\right]^2,
\end{equation}
where $W=100~\mathrm{s} $ is a velocity weight that converts the velocity residual to an equivalent distance scale. Including both residuals prevents a satellite that passes through a similar position with an inconsistent velocity from obtaining a high ranking. For every catalogue object, $J$ was minimised independently over the allowed time interval, and the resulting minimum values were used to construct the final catalogue ranking. The robustness of the identification to the choice of $W$ is examined in Section~\ref{subsect:robustness}.

The blind search returned STARLINK-32838 (NORAD ID 62827) as the clearly preferred candidate. Its TLE had an epoch of 2025-04-13 18:00:02 UTC, approximately 44~h before the Event~1 nominal epoch. The best match occurred at an offset of $-28.302$~s and gave a position residual of 5.47~km and a velocity residual of 13.42~$\mathrm{m\,s^{-1}}$, with a final score of $31.7~\mathrm{km^2}$. The second-ranked object, STARLINK-31335, had substantially larger residuals of 186.85~km and 296.61~$\mathrm{m\,s^{-1}}$, with a score of $3.58\times10^4~\mathrm{km^2}$.

\begin{figure*}
        \centering
        \includegraphics[width=\hsize]{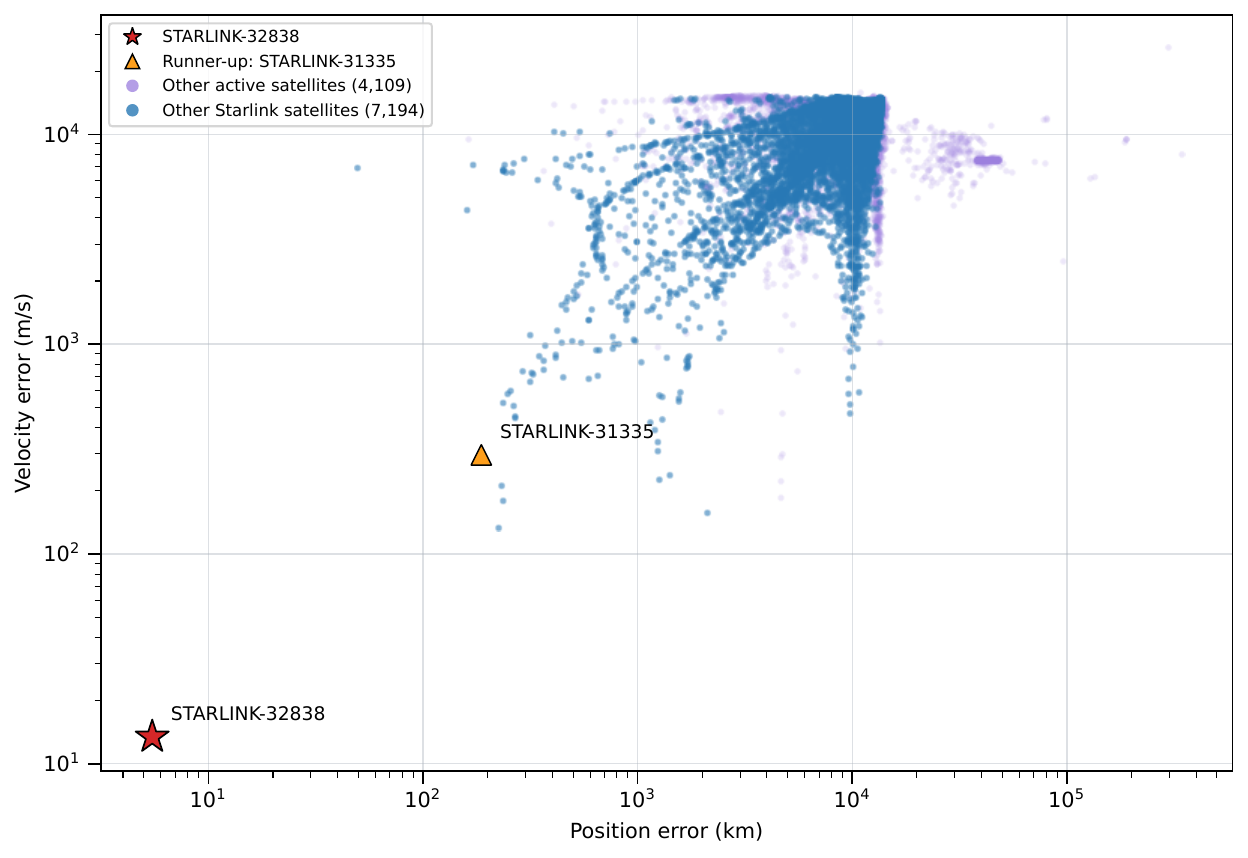}
        \caption{
Position and velocity residuals between the Event~1 payload state vector and the TLE predicted states of all 11\,305 active satellites. For each object, the time offset was independently optimised within $\pm180$~s by minimising the joint matching score. The red five-pointed star marks the best-matching candidate, STARLINK-32838, while the orange triangle marks the runner-up, STARLINK-31335. Other Starlink satellites and other active satellites are shown as blue and light-purple circles, respectively. Both axes are logarithmically scaled.
}
        \label{pic:catalogue_match}
    \end{figure*}

As shown in Figure~\ref{pic:catalogue_match}, STARLINK-32838 (five-pointed star) is clearly separated from the remaining catalogue objects in the position--velocity error plane. The $-28.302$~s offset is the time displacement that best aligns the earlier TLE prediction with the payload state and should not be interpreted as a revised reception time. The full-catalogue separation provides strong independent support for identifying the Event~1 transmitter as STARLINK-32838.

\subsection{Robustness to the velocity weight}
\label{subsect:robustness}

The velocity weight $W$ controls the relative contribution of the velocity residual to the joint matching score and is a manually specified parameter. To determine whether the ranking depended on its adopted value, we repeated the complete full-catalogue ranking for seven analysis weights:
$$
W \in \{0,10,30,100,300,1000,3000\}~{\mathrm{s}}.
$$
The case $W=0$ corresponds to position-only matching, while progressively larger values assign increasing importance to the velocity residual. For each value of $W$, the time offset was re-optimised independently over the same $\pm180$~s interval for all 11\,305 catalogue objects.

\begin{figure*}
        \centering
        \includegraphics[width=\hsize]{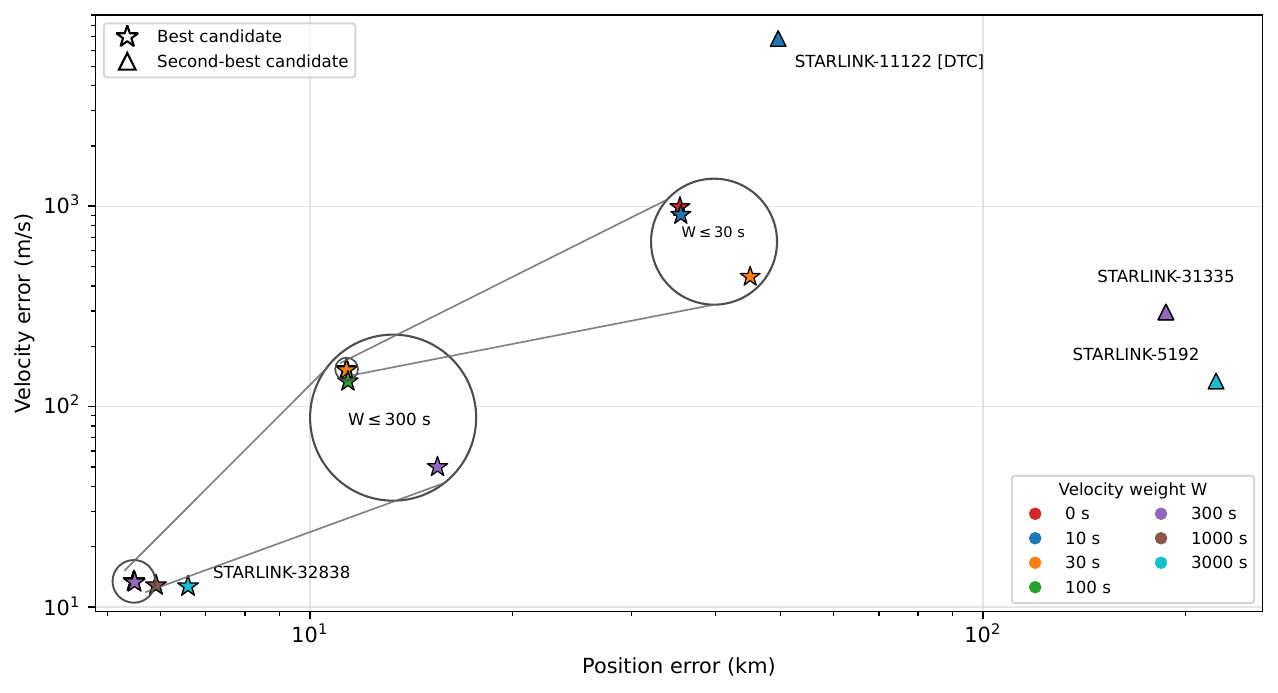}
        \caption{Dependence of the two highest-ranked candidates on the velocity weight $W$. Both main axes are logarithmically scaled. Five-pointed stars denote the highest-ranked candidate and triangles denote the runner-up, with colours indicating the adopted value of $W$. STARLINK-32838 remains the highest-ranked candidate in every case, whereas the runner-up changes among three satellites. The circular callouts provide successive magnifications of the overlapping STARLINK-32838 markers. Several runner-up markers overlap because their residual differences are too small to be resolved on the logarithmic axes or because their optimised time offsets reach the boundary of the search interval.}
        \label{pic:robustness}
    \end{figure*}

As shown in Figure~\ref{pic:robustness}, STARLINK-32838 remained the highest-ranked candidate for every tested velocity weight, with the seven star markers forming a compact cluster at the lower left of the position--velocity error plane. The second-ranked candidate changed from STARLINK-11122 [DTC] to STARLINK-31335 and then to STARLINK-5192 as $W$ increased. Nevertheless, the ratio of the runner-up score to the STARLINK-32838 score never fell below 82. The identification is therefore insensitive to the precise balance assigned to the position and velocity residuals. We retained $W=100$~s as the fiducial velocity weight for the subsequent TLE-based Doppler comparison.

\subsection{Association of Event~2 through differential Doppler}
\label{subsect:doppler}

The state-vector fields of Event~2 are zero and therefore cannot be used for an independent blind catalogue search. Its consecutive message number, identical internal protocol identifier, and 100~s increment in the UTC timecode already suggest that the two packets originated from the same source. We further tested this association by comparing the measured change in carrier-frequency offset (CFO) with the Doppler evolution predicted for STARLINK-32838. An absolute CFO may contain contributions from the satellite motion, transmitter frequency offset, receiver sampling clock, and uncertainty in the nominal channel centre. Because these instrumental and transmitter contributions should vary little over the approximately 100~s interval between the events, taking the CFO difference suppresses most of the common offset.

For this comparison, the CFO of each event was estimated independently from all 15 ordinary preamble upchirps, with their median adopted as the value. A conservative empirical uncertainty was estimated by combining the chirp-to-chirp sample standard deviation with an FFT-grid quantisation term in quadrature. The uncertainties of the two events were then combined in quadrature to obtain the uncertainty in their CFO difference. This gives
$$
\Delta f_{\mathrm{CFO}}
=f_{{\mathrm{CFO}},2}-f_{{\mathrm{CFO}},1}
=-3499.35\pm14.43~{\mathrm{Hz}},
$$
where $f_{{\mathrm{CFO}},i}$ denotes the median CFO estimated from the 15 preamble upchirps of Event $i$. The negative value indicates that the received frequency of Event~2 was lower than that of Event~1.

\begin{figure*}[t]
        \centering
        \includegraphics[width=\hsize]{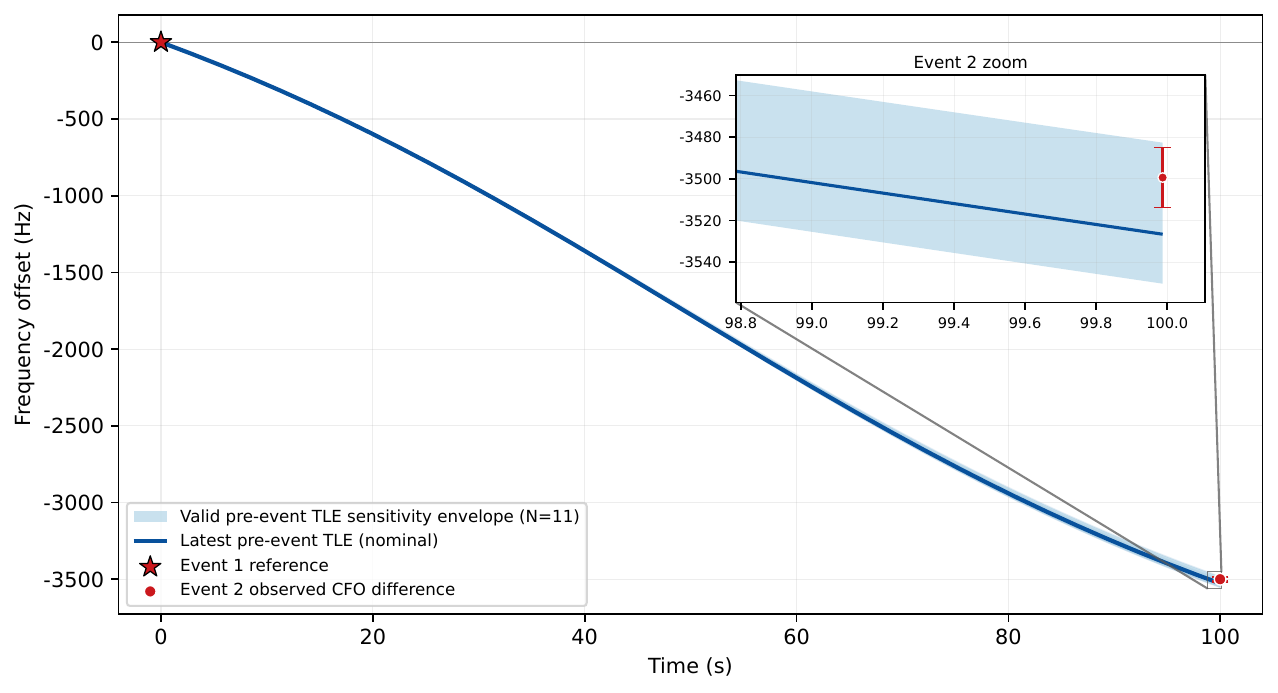}
        \caption{Differential Doppler comparison between Event~1 and Event~2 for 
        STARLINK-32838. Both axes are referenced to Event~1: $t=0$ corresponds to its preamble start, and the frequency offset denotes the change relative to its frequency. Event~2 occurs 99.987~s later. The red five-pointed star marks the Event~1 reference and the red circle shows the measured Event~2 CFO difference; its vertical error bar represents the conservative empirical uncertainty. The blue curve gives the TLE-predicted differential Doppler evolution, and the shaded envelope describes its sensitivity to the choice of pre-event TLE rather than a statistical confidence interval.
            }
        \label{pic:doppler}
    \end{figure*}

The predicted Doppler shift was calculated from the line-of-sight range rate between the E08 pod and STARLINK-32838 at 137.055~MHz. The GPS-derived epoch of Event~1 served as the reference for calculating the predicted Doppler difference over the observed 99.987~s preamble interval, with the same best-fit TLE time offset applied to both events. Historical TLE records for STARLINK-32838 were obtained from Space-Track \footnote{\url{https://www.space-track.org/}}. We considered the 12 available TLEs with epochs between 10 April 2025 and Event~1, each independently aligned with the Event~1 state vector using the fiducial velocity weight $W=100$~s. After excluding one poorly matched TLE, the shaded sensitivity envelope in Figure~\ref{pic:doppler} is constructed from the pointwise minima and maxima of the 11 retained differential Doppler curves. The latest pre-event TLE, with an epoch of 13 April 2025 18:00:02 UTC, provides the nominal prediction shown by the dark-blue curve. At Event~2, the nominal TLE predicts a frequency change of $-3526.55$~Hz, giving an observed-minus-predicted residual of $+27.20$~Hz. As highlighted in the enlarged view, the measured CFO difference and its conservative empirical uncertainty interval lie within the $-3550.30$ to $-3482.59$~Hz range spanned by the TLEs. This agreement strongly supports the conclusion that Event~2 was also transmitted by STARLINK-32838.

\subsection{Post-identification validation using SpaceX-E SupGP data}

Following the blind identification of STARLINK-32838, we performed a post-identification validation using historical SpaceX-E supplemental general perturbations (SupGP) data obtained from CelesTrak. These elements were derived from operator-provided ephemerides and fitted for propagation with SGP4. The nearest SupGP epoch preceding Event~1 was 15 April 2025 09:35:42 UTC, approximately 4.4~h before the GPS-derived epoch.

Applying the same $W=100$~s joint score optimisation used in Section~\ref{subsect:blind_comparison} gave a best-fitting time offset of only $ \Delta t_{\mathrm{SupGP}}=+0.01995~{\mathrm{s}}$, with position and velocity residuals of $0.288$~km and $5.29~\mathrm{m\,s^{-1}}$. The SupGP data were also used to predict the differential Doppler shift over the same 99.987~s interval, using the same fiducial weight $W=100~\mathrm{s}$. The predicted differential Doppler shift is $\Delta f_{\mathrm{SupGP}}=-3528.48~{\mathrm{Hz}}$. 

Compared with the nearest TLE, the SupGP data produced substantially smaller residuals with an almost negligible time adjustment. This improvement is consistent with the expectation that orbital data closer in time to the event should provide a more accurate representation of the satellite state. The SupGP validation therefore provides stronger support for both the identification of STARLINK-32838 and the inferred interpretation of the payload state fields.

\section{Orbital context}
\label{section4}

Following the satellite identification, we examined the orbital context of the detected VHF TT\&C packets from three complementary perspectives. First, we used the measured event interval to identify cadence-predicted transmission opportunities within the five observing windows and evaluated the visibility of STARLINK-32838 from the 21CMA at each epoch. Second, we examined the short- and long-term orbital evolution of STARLINK-32838 to determine its orbital state around the two events. Finally, we reanalysed the public EDA2 catalogue and the corresponding historical TLE records to investigate the relationship between these emissions and satellite orbital state. These analyses show that the detection of only two packets is consistent with the observing geometry. Both packets occurred while STARLINK-32838 was undergoing orbit raising; however, the broader EDA2 sample indicates that 137.05~MHz emission is not confined to this orbital phase.

\subsection{Orbital visibility}

Assuming that the satellite transmitted periodically throughout the five observing windows at the measured cadence, candidate transmission epochs were identified at integer multiples of 99.987~s relative to the Event~1 preamble start. This yielded seven, six, nine, two, and two candidate epochs within W1--W5, respectively, giving 26 in total. For each candidate opportunity, the orbital geometry of STARLINK-32838 was evaluated using the SupGP data whose epoch was closest to the predicted time. To account for possible cumulative timing errors introduced by the extrapolation, the azimuth, elevation, and slant range relative to the 21CMA were calculated at the predicted epoch and at 1~s on either side. An opportunity was considered geometrically visible only if all three sampled elevations were above the horizon.

\begin{figure}
        \centering
        \includegraphics[width=\hsize]{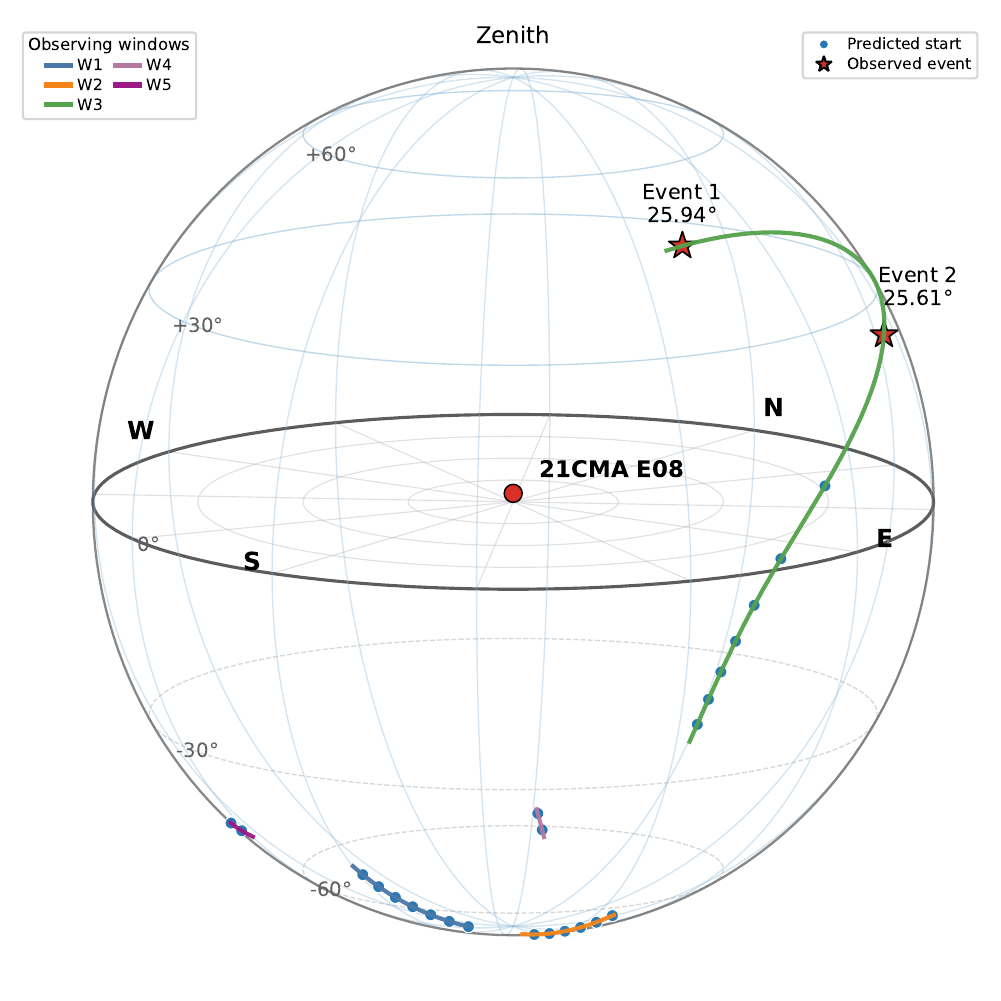}
        \caption{Local ENU unit-sphere representation of the apparent trajectory of STARLINK-32838 relative to the 21CMA E08 during the five observing windows. All positions are projected onto a unit sphere to represent direction only; radial distance from the centre does not encode the satellite slant range.
        }
        \label{pic:local_sky}
    \end{figure}

As shown in Figure~\ref{pic:local_sky}, we constructed a local east--north--up (ENU) sky representation centred on the E08, showing the trajectory of STARLINK-32838 during each of the five observing intervals. The unit-sphere projection represents direction only, rather than satellite range. The blue points mark all cadence-predicted transmission opportunities, while the two detected packets are indicated by red five-pointed stars. Of the 26 predicted opportunities, 22 occurred below the horizon. Only four consecutive opportunities were geometrically visible, all during observation W3, with the first two corresponding to Event~1 and Event~2. At their preamble start times, Event~1 and Event~2 had elevations of $25.94^\circ$ and $25.61^\circ$ and slant ranges of 671~km and 677~km, respectively. The following two opportunities occurred at substantially lower elevations of $9.85^\circ$ and $1.41^\circ$ and greater slant ranges of 1226~km and 1891~km. Under inverse-square propagation, their greater ranges correspond to additional attenuation of approximately 5.16 and 8.92~dB relative to Event~2, the weaker of the two detected events. The 21CMA is located in a valley surrounded by mountainous terrain. The lower elevations and greater slant ranges of the latter two opportunities, together with the potential effects of terrain obstruction and the sidelobe response, may have reduced their detectability. This interpretation, however, assumes that STARLINK-32838 continued transmitting VHF TT\&C packets at the inferred intervals.

\begin{figure*}
        \centering
        \includegraphics[width=\hsize]{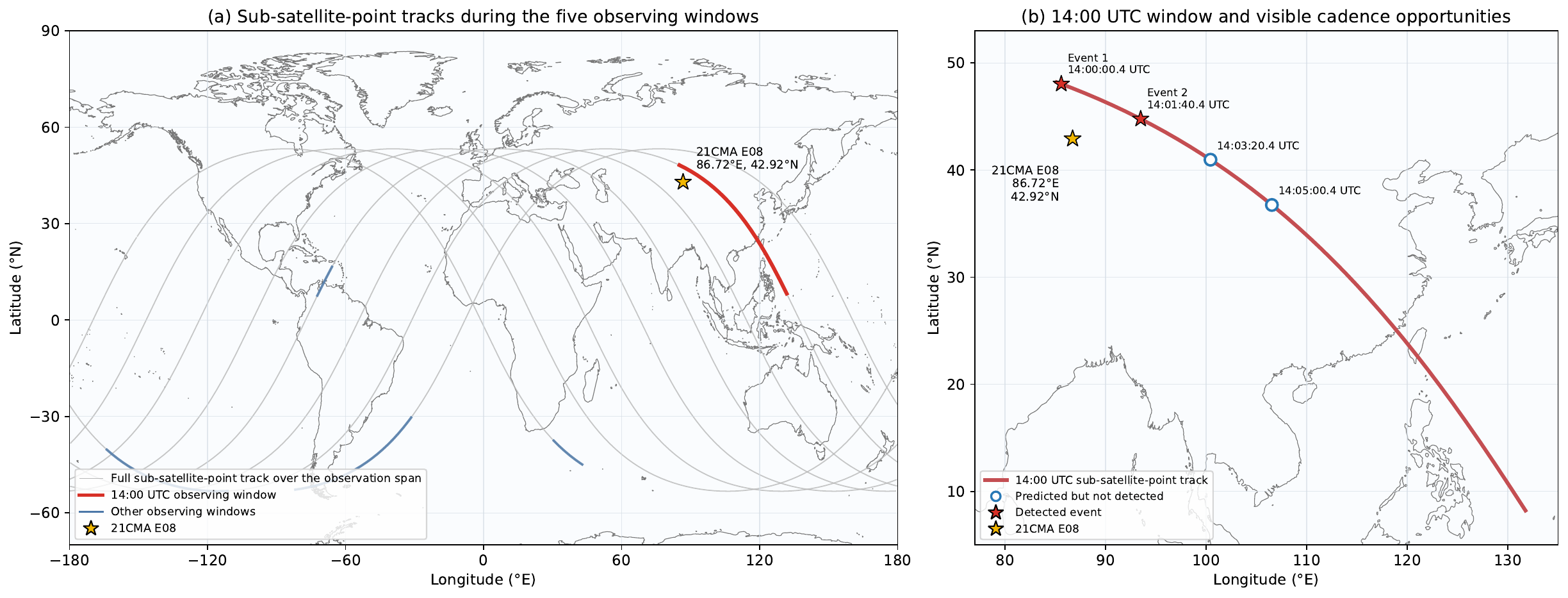}
        \caption{Sub-satellite-point trajectories of STARLINK-32838 during the 21CMA observations.}
    
        \label{pic:subsatellite}
    \end{figure*}

Figure~\ref{pic:subsatellite} provides the corresponding geographical context. The left panel shows the propagated sub-satellite-point track of STARLINK-32838 over the full time span covered by the five observations, with the track segments sampled during the observing windows highlighted. Among these intervals, only the W3 track passed close to 21CMA. The right panel enlarges this passage and marks the four geometrically visible cadence-predicted opportunities. From Event~1 to the two subsequent undetected opportunities, the sub-satellite point moved progressively eastward and southward away from the 21CMA, consistent with the decrease in elevation and increase in slant range inferred from the topocentric analysis.

\subsection{Orbital evolution}

The public TLE history of STARLINK-32838 from Space-Track is discontinuous around the time of the two events. The last TLE preceding Event~1 has an epoch of 13 April 2025 18:00 UTC, whereas the next available TLE has an epoch of 17 April 2025 21:38 UTC, leaving a gap of 99.64~h that contains both events. However, the SpaceX-E SupGP data provide continuous coverage across this interval and were therefore used to examine the short-term orbital evolution. 

\begin{figure*}
        \centering
        \includegraphics[width=\hsize]{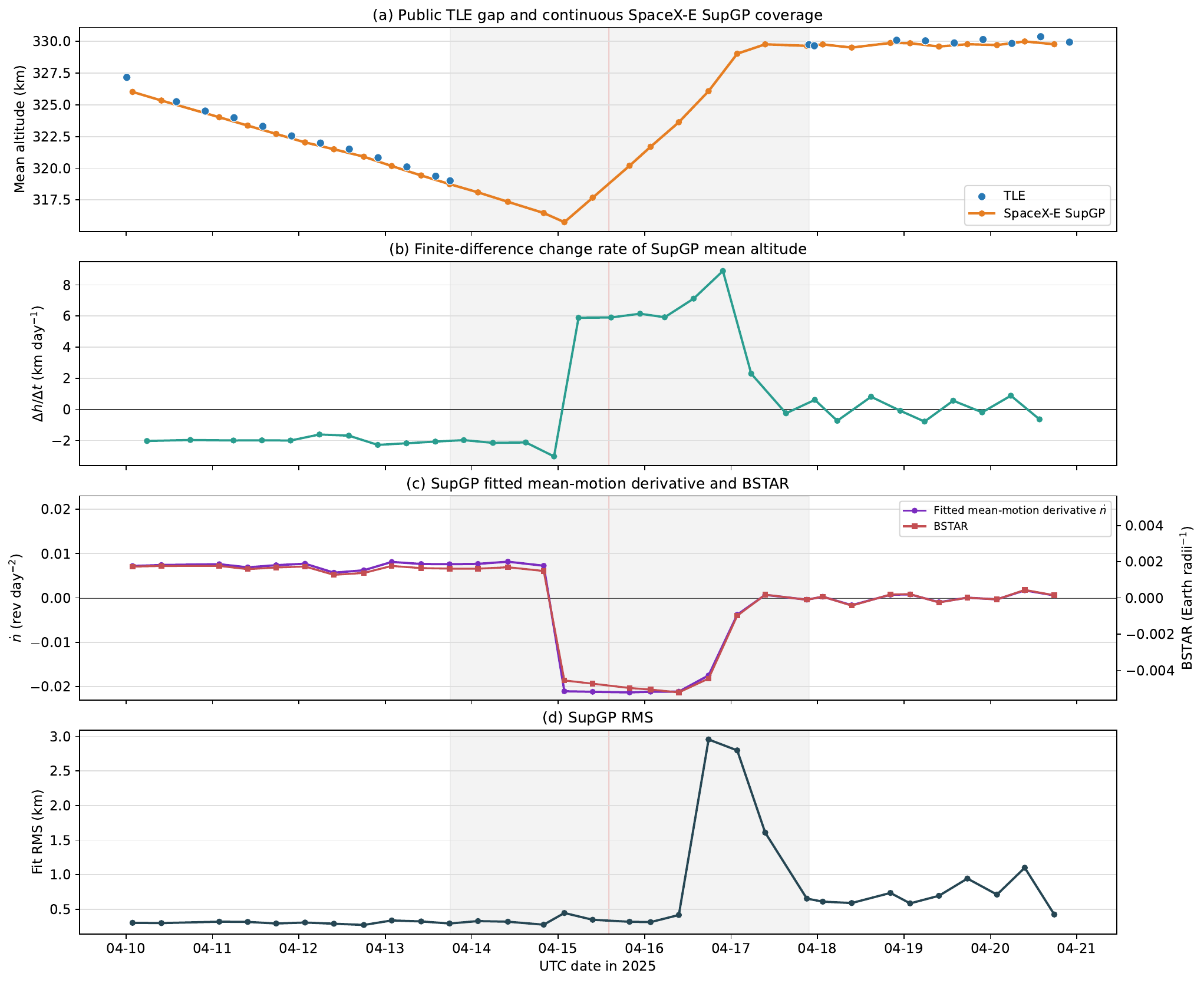}
        \caption{Short-term orbital evolution of STARLINK-32838 around the two events. In panels~(a), (c), and (d), each marker is plotted at the epoch of the corresponding TLE or SupGP entry; in panel~(b), each marker represents the finite difference between two adjacent SupGP entries and is plotted at their temporal midpoint. The grey shaded region denotes the 99.64~h gap between the TLEs bracketing the events, while the narrow red band marks the interval between Event~1 and Event~2. The lines joining successive markers illustrate the temporal trends and do not represent continuous orbital solutions between the sampled epochs.
       }
        \label{pic:evolution}
    \end{figure*}

In all four panels of Figure~\ref{pic:evolution}, the grey shaded region marks the 99.64~h TLE gap. The narrow red band marks the interval between Event~1 and Event~2; because the events are separated by only 99.987~s, the band appears as a single vertical line on this timescale. 

As shown in Figure~\ref{pic:evolution}(a), the mean orbital altitude inferred from the SupGP mean motion decreased until early on 15 April, reaching approximately 315.7~km, and subsequently began to increase. The finite difference between the SupGP entries immediately preceding and following Event~1 indicates that the mean orbital altitude was increasing at approximately 5.9~km~day$^{-1}$, as shown in Figure~\ref{pic:evolution}(b). Both detected packets therefore occurred during the rising portion of this orbital evolution. Figure~\ref{pic:evolution}(c) shows the fitted mean-motion derivative, which describes the rate of change in orbital mean motion, together with the SGP4 BSTAR parameter, which primarily represents the fitted drag-like contribution. Both parameters changed markedly as the altitude began to rise, and their evolution is consistent with the fitted-parameter behaviour expected during orbit raising. Figure~\ref{pic:evolution}(d) shows the RMS difference between each SupGP fit and the source operator ephemeris. The RMS increased to approximately 3~km during the rapid rise. This temporary degradation in the SGP4 fit, together with the altitude and fitted-parameter changes, supports the interpretation that STARLINK-32838 was undergoing active orbit raising. The occurrence of the packets during the raising episode is consistent with the authorised use of the VHF beacon as a backup TT\&C system during launch and early-orbit operations \citep{fcc2024starlinkbeacon}. This temporal correspondence alone, however, does not imply that the 137.055~MHz transmissions are confined to orbit raising.

We next compared STARLINK-32838 with the other 22 satellites from the same Group 11-6 launch. For each spacecraft, the TLE with an epoch closest to the Event~1 preamble start was selected. The other group members occupied a narrow range of mean orbital altitudes between 447.24 and 447.58~km. By contrast, the nearest TLE for STARLINK-32838 gave a mean altitude of 319.01~km, making it a clear orbital outlier within the launch group.

\begin{figure}
        \centering
        \includegraphics[width=\hsize]{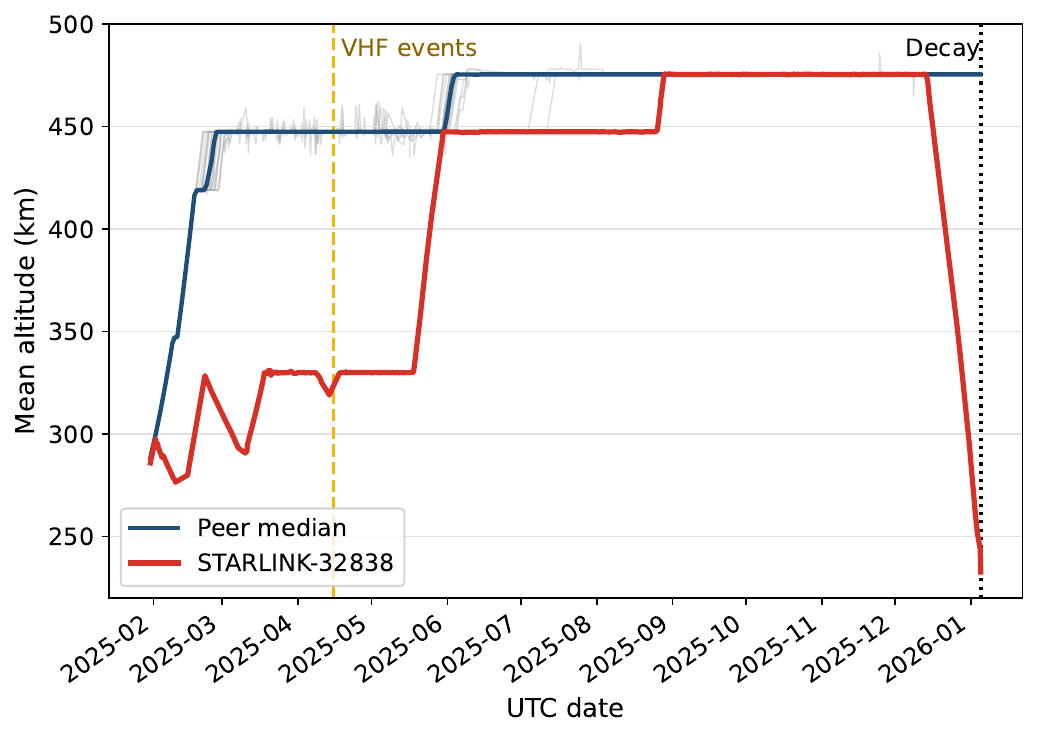}
        \caption{Long-term TLE-derived mean-altitude evolution of STARLINK-32838 and the other spacecraft in Starlink Group 11-6. The grey curves show the individual histories of the other 22 spacecraft, the blue curve shows their daily median altitude, and the red curve shows STARLINK-32838. The yellow dashed line marks the two VHF events and the black dotted line marks the decay date of STARLINK-32838.
       }
    
        \label{pic:longevolution}
    \end{figure}

The longer-term TLE history of these satellites provides additional context for this unusual state. As shown in Figure~\ref{pic:longevolution}, while most of the group reached an altitude near 447~km shortly after launch and subsequently followed similar trajectories, STARLINK-32838 remained between approximately 280 and 330~km during its first several months. The two events occurred while the spacecraft was still far below the group orbit and had begun the short-term altitude increase. STARLINK-32838 subsequently entered a pronounced orbit-raising phase and eventually reached the altitude of the other satellites, before being catalogued as decayed on 5 January 2026. The long-term history therefore also supports the conclusion that STARLINK-32838 was an orbital outlier within Group 11-6.

\subsection{Orbital states of previous EDA2 detections}

The closest observational counterparts to our two events are the periodic Starlink IEMR previously detected with the EDA2. \citet{EDA2_2023} detected 1,130 flashes from 136 Starlink satellites, subsequently classified as v1.5 and v2-mini Ku spacecraft, within a 0.926~MHz observing band centred at 137.5~MHz. Most flashes recurred approximately every 100~s and lasted less than 2~s. SpaceX engineers confirmed that Starlink satellites carried 'Swarm' transmitters operating near 137~MHz for telemetry, tracking, and control, leading the authors to interpret the flashes as intended TT\&C transmissions. Using observations centred at 136.72~MHz with an effective bandwidth of approximately 0.90~MHz, \citet{EDA2_2025} subsequently detected similar 100~s periodic narrowband emission from both v1.5 and v2-mini Ku satellites and constrained its frequency to approximately 137.05~MHz. 

The frequencies, cadence, and duration constraint are all consistent with the packets detected here at 137.055~MHz. The correspondence also extends to the spacecraft model: STARLINK-32838 is a v2-mini Ku satellite. Our results therefore provide further evidence that the emissions reported in the two EDA2 studies and the events here belong to the same class of Starlink VHF TT\&C transmissions. Moreover, our observations provide a more detailed packet-level analysis and extend the geographical scope of its detection from Australia to China.

To determine whether the occurrence of 137.05~MHz emission during orbit raising, as observed for STARLINK-32838, is representative of the broader emitting population, we reanalysed the public EDA2 catalogue \footnote{\url{https://zenodo.org/records/15089853}} following the fine-channel Gaussian-classification procedure of \citet{EDA2_2025}. After selecting the narrowband emission near 137.05~MHz and combining detections belonging to the same emission episode, we obtained 75 episodes associated with 26 satellites. We then examined their historical TLE records obtained from Space-Track to determine the orbital state of each satellite around its detection date.

The 75 emission episodes were grouped by satellite and UTC date, yielding 28 satellite--date combinations. For each TLE epoch, the mean orbital altitude was derived from the semimajor axis, and the median value on each UTC date was used to construct a daily altitude series. The altitude-change rate, denoted by $\dot{h}_{14}$, was estimated within $\pm14$~days of the detection date using the Theil--Sen estimator, which takes the median of the altitude-change rates calculated between every pair of daily data points.

Orbital states were classified from the altitude-change rate. A satellite--date combination was classified as locally stable when $|\dot{h}_{14}|\leq0.2~\mathrm{km\,d^{-1}}$, as orbit raising when $\dot{h}_{14}>0.2~\mathrm{km\,d^{-1}}$, and as orbit lowering when $\dot{h}_{14} <-0.2~\mathrm{km\,d^{-1}}$. The classifications were additionally cross-checked against the longer-term altitude histories to distinguish a stable orbital phase from a temporary pause during orbit raising or lowering . The complete per-satellite altitude histories are available here \footnote{\url{https://github.com/JinYi1108/21CMA_Starlink_VHF/tree/main/paper_results/orbit_phase_panels}}.

Of the 28 satellite--date combinations, 17 occurred during orbit raising, nine during locally stable orbital phases, and two during orbit lowering. The nine stable combinations represented seven distinct satellites. One of these cases, STARLINK-30262 (NORAD ID 57535), was detected during a temporary altitude plateau within a longer-term orbital-decay trajectory: its altitude had already begun to decrease before the detection, remained locally stable around the detection date, and subsequently resumed decreasing until catalogue decay. The classification therefore describes the orbital behaviour around the detection date rather than the subsequent long-term evolution of the satellite. The derived orbital properties and classifications of all 28 satellite--date combinations are listed in Table~\ref{table:eda2_orbital_states}.

These results show that the 137.05~MHz signal is not observationally confined to the launch and orbit-raising phases; it can also be detected while satellites are in locally stable orbital phases or undergoing orbit lowering. The FCC order authorises the VHF beacons as a backup TT\&C system both during orbit raising and in emergency situations at authorised operational altitudes, providing one possible operational context for some detections outside the orbit-raising phase \citep{fcc2024starlinkbeacon}. However, the orbital records alone cannot establish whether these transmissions were associated with emergency use or with other operational circumstances. Additional detections, together with more detailed orbital and operational information, will be required to determine when and why the VHF beacon is activated outside the orbit-raising phase.

\begin{table*}
\caption{Orbital states of the EDA2 satellite--date detections at 137.05~MHz.}
\label{table:eda2_orbital_states}
\centering
\small
\begin{tabular}{lrrrll}
\hline\hline
Satellite &
NORAD ID &
Date (UTC) &
$\bar{h}_{\mathrm{det}}$ (km) &
$\dot{h}_{14}$ (km d$^{-1}$) &
Orbital state \\
\hline
STARLINK-5537  & 55384 & 2024-08-15 & 559.0 & 0.00  & Locally stable \\
STARLINK-5556  & 55385 & 2024-08-15 & 559.0 & 0.00  & Locally stable \\
STARLINK-5329  & 56513 & 2024-08-15 & 559.0 & 0.00 & Locally stable \\
STARLINK-5336  & 57138 & 2024-08-15 & 559.1 & 0.00  & Locally stable \\
STARLINK-30495 & 57968 & 2024-08-15 & 559.1 & 0.00 & Locally stable \\
STARLINK-32128 & 60272 & 2024-08-15 & 386.8 & 7.06 & Orbit raising \\
STARLINK-31745 & 60329 & 2024-08-15 & 316.6 & 3.48 & Orbit raising \\
STARLINK-32077 & 60344 & 2024-08-15 & 314.7 & 3.78 & Orbit raising \\
STARLINK-32186 & 60412 & 2024-08-15 & 289.7 & 3.87 & Orbit raising \\
STARLINK-32265 & 60415 & 2024-08-15 & 288.9 & 3.97 & Orbit raising \\
STARLINK-30262 & 57535 & 2024-08-16 & 444.0 & 0.00 & Locally stable \\
STARLINK-31921 & 59868 & 2024-08-16 & 327.0 & 2.34 & Orbit raising \\
STARLINK-11158 & 60403 & 2024-08-16 & 299.2 & 3.40 & Orbit raising \\
STARLINK-5537  & 55384 & 2024-10-23 & 559.1 & 0.00 & Locally stable \\
STARLINK-5556  & 55385 & 2024-10-23 & 559.1 & 0.00 & Locally stable \\
STARLINK-5095  & 56885 & 2024-10-23 & 559.1 & 0.00 & Locally stable \\
STARLINK-30406 & 57881 & 2024-10-23 & 461.2 & $-0.36$ & Orbit lowering \\
STARLINK-31703 & 59861 & 2024-10-23 & 305.6 & $-3.14$ & Orbit lowering \\
STARLINK-32406 & 61545 & 2024-10-23 & 304.1 & 7.76 & Orbit raising \\
STARLINK-32411 & 61547 & 2024-10-23 & 304.2 & 7.75 & Orbit raising \\
STARLINK-32389 & 61549 & 2024-10-23 & 303.6 & 7.87 & Orbit raising \\
STARLINK-32366 & 61551 & 2024-10-23 & 303.7 & 7.85 & Orbit raising \\
STARLINK-32325 & 61588 & 2024-10-23 & 300.5 & 5.84 & Orbit raising \\
STARLINK-32416 & 61589 & 2024-10-23 & -- & 6.06 & Orbit raising \\
STARLINK-32413 & 61590 & 2024-10-23 & -- & 6.06 & Orbit raising \\
STARLINK-32409 & 61591 & 2024-10-23 & 299.2 & 5.96 & Orbit raising \\
STARLINK-32422 & 61592 & 2024-10-23 & 298.9 & 6.12 & Orbit raising \\
STARLINK-32397 & 61593 & 2024-10-23 & 298.6 & 6.09 & Orbit raising \\
\hline
\end{tabular}

\tablefoot{
For each satellite--date combination, the representative detection epoch was defined as the midpoint between the start of its first detected emission episode and the end of its last. The mean orbital altitude $\bar{h}_{\mathrm{det}}$ was linearly interpolated between the TLE
records bracketing this epoch. The altitude-change rate $\dot{h}_{14}$ is the Theil--Sen slope of the daily median mean orbital altitude within a $\pm14$-day window centred on the detection date. NORAD~61589 and 61590 had only recently been launched when detected, and no pre-detection TLE records were available. Their mean altitudes could therefore not be interpolated, and both trend estimates were derived solely from the available post-detection records.}
\end{table*}

\section{Flux density}
\label{section5}
In this section, we estimated the incident flux density of the detected emission. Because both packets were received through sidelobes, converting the measured power into flux density requires accounting for the direction-dependent station response. We combined the on--off power measurements with the estimated system equivalent flux density (SEFD) and an idealised OSKAR station-beam model. Given the systematic uncertainty of the modelled sidelobe response, the resulting flux densities are interpreted as order-of-magnitude estimates.

Event~1 provides the more reliable estimate, yielding approximately 20.6~MJy at a slant range of 669.5~km, equivalent to 9.23~MJy at 1000~km and an equivalent isotropic radiated power (EIRP) of approximately 48.3~mW. Event~2 was not used for this estimate because it lies in a very low-response region of the idealised beam model. We further compared Event~1 with the previous EDA2 detections near 137~MHz. After approximate normalisation for their different observing configurations, all three measurements are of order 10~MJy at 1000~km.

\subsection{Flux density estimation}

As the first step in estimating the flux densities of the two events, we obtained the mean on-event power, $P_{\mathrm{on}}$, by integrating the linear power spectral density over a 41.667~kHz bandwidth centred on each event's measured carrier frequency and averaging over the 1.027584~s packet duration. The background power, $P_{\mathrm{off}}$, was estimated from two 1~s intervals before and after each packet, each separated from the packet by a 0.30~s guard interval. The fractional excess power, $\eta=P_{\mathrm{on}}/P_{\mathrm{off}}-1$, was 35.663 and 1.022 for Event~1 and Event~2, respectively.

Using the SEFD relation given in Eq.~(1) of our previous work \citep{21CMA_UEMR}, together with the same E08 effective area and non-sky temperature contribution adopted at 150~MHz, the sky temperature contribution was scaled to 137.055~MHz following the $T_{\mathrm{sky}}\propto\nu^{-2.55}$ dependence. This yielded a system temperature of 415.5~K and a main-beam SEFD of 5263~Jy. The nominal main-beam-equivalent flux density was then calculated as
\begin{equation}
    S_{\mathrm{eq,MB}}=\eta\,\mathrm{SEFD}_{\mathrm{MB}},
\end{equation}
giving 187.7~kJy and 5.38~kJy for Event~1 and Event~2, respectively, before accounting for sidelobe response along the satellite track during each packet.

The E08 response in the actual satellite directions was estimated using an idealised station-beam model implemented with OSKAR \footnote{\url{https://github.com/OxfordSKA/OSKAR}} \citep{OSKAR}. The model incorporated the positions of all 127 antenna elements, assuming identical isotropic elements and uniform weights, with the NCP adopted as the phase centre at 137.055~MHz. The resulting power response was normalised to unity in the phase-centre direction. The SupGP orbit was propagated throughout each packet to determine the satellite direction. The relative power response was evaluated at approximately 5~ms intervals and averaged in linear units. The resulting relative gains were $-20.40$~dB for Event~1 and $-47.07$~dB for Event~2. Because the model does not include mutual coupling, polarisation mismatch, instrumental gain and phase errors, these values represent model-dependent array-factor estimates rather than calibrated sidelobe gains.

Event~2 lies in a very low-response region of the idealised beam model, where even modest modelling errors in the sidelobe response would produce large changes in the inferred flux density. We therefore used only Event~1, which occurred in a substantially higher-response direction, to obtain an order-of-magnitude estimate of the incident flux density. It was calculated as
\begin{equation}
    S_{\nu}
    = \frac{S_{\mathrm{eq,MB}}}{\overline{G}_{\mathrm{rel}}},
\end{equation}
yielding approximately 20.6~MJy for Event~1 at a packet-averaged slant range of 669.5~km. After scaling only for inverse-square propagation, this corresponds to approximately 9.23~MJy at a reference distance of 1000~km and to an EIRP of approximately 48.3~mW with a 41.667~kHz occupied bandwidth.

\subsection{Comparison with previous EDA2 observations at 137~MHz}

For an approximate flux-density comparison, \citet{EDA2_2023} reported a maximum value of approximately 1~MJy~beam$^{-1}$ at a slant range of about 500~km. This value was obtained by averaging over 4~s and the full 0.926~MHz observing bandwidth and by summing the beam-corrected XX and YY polarisations. Because the apparent angular size of a Starlink satellite is far smaller than the angular resolution of the EDA2 images, the peak intensity in Jy~beam$^{-1}$ can therefore be treated as the source flux density for this comparison. After normalising to 1000~km and assuming that these flashes had the same 1.027584~s duration and 41.667~kHz occupied bandwidth as Event~1, the reported maximum corresponds to approximately 21.6~MJy. If the received power were divided equally between the two linear polarisations, the corresponding single-polarisation value would be approximately 10.8~MJy, comparable to our estimate of 9.23~MJy at 1000~km. This agreement should be interpreted only at the order-of-magnitude level because the actual duration, occupied bandwidth, and polarisation of the flashes were not resolved.

The finer spectral resolution of \citet{EDA2_2025} enables a more direct comparison. In the strongest 137.05~MHz record in their public catalogue, the two adjacent fine channels containing the emission, each 28.9375~kHz wide, reached 102.85 and 97.07~MJy~beam$^{-1}$ in the YY polarisation at a slant range of 430.4~km. Their mean corresponds to a source flux density of approximately 99.96~MJy. Normalising only for distance gives approximately 18.5~MJy at 1000~km. If the emission is additionally assumed to have the same duration as Event~1, correcting for dilution within the 2~s EDA2 integration increases this estimate to approximately 36.0~MJy. This value is about 5.9~dB higher than our Event~1 estimate of 9.23~MJy, but the two measurements remain consistent at the order-of-magnitude level. The comparison remains subject to uncertainty in the EDA2 packet duration and its alignment within the 2~s integration, as well as differences in polarisation and beam-response calibration.

Despite the different observing and averaging schemes, the approximately normalised EDA2 measurements and our Event~1 estimate are all of order 10~MJy at 1000~km, providing an independent order-of-magnitude consistency check on the signal strength.

\section{Discussion}
\label{section6}

Previous observations have established the presence of recurrent Starlink emissions near 137~MHz \citep{EDA2_2023,EDA2_2025}. In this work, information recovered through signal demodulation and decoding enabled the identification of STARLINK-32838 through comparison with orbital data. This demonstrates that decoded signal content can complement conventional spectral and orbital methods for identifying satellite-associated RFI. Nevertheless, the proposed interpretation of the payload state-vector fields remains to be verified using additional packets from different spacecraft. More broadly, these results demonstrate the capability of the 21CMA to detect and characterise satellite VHF emissions relevant to low-frequency radio astronomy. The detection of these emissions in China, where Starlink does not provide commercial broadband service, further shows that the absence of local service does not preclude satellite emissions capable of affecting local radio astronomical observations.

The two detected packets have the same duration of 1.027584~s and are separated by approximately 100~s, but this limited sample cannot establish whether either property remains stable across different satellites or operational states. The visibility analysis is therefore conditional on the inferred transmission cadence and does not confirm that transmissions occurred at every predicted epoch. Longer observations covering multiple passes and spacecraft are needed to characterise the variability of packet durations and transmission intervals.

The main limitation of the flux density estimate is the uncertainty in the E08 sidelobe response. Although the excess power was measured directly, its conversion to incident flux density relies on an idealised OSKAR beam model that does not fully account for element patterns, mutual coupling, polarisation effects, or instrumental imperfections. Consequently, the Event~1 flux density should be regarded as an order-of-magnitude estimate. The approximate agreement with previous EDA2 measurements provides a useful consistency check but does not independently validate the adopted beam model. Improved beam modelling and empirical calibration would help constrain these uncertainties.

The detection of STARLINK-32838 during orbit raising is consistent with the FCC-authorised role of the VHF beacon as a backup TT\&C system during launch and early-orbit operations. However, our re-analysis of the data from \citet{EDA2_2025} reveals that similar emissions also occurred during locally stable orbital phases and orbit lowering, indicating that VHF activity is not restricted to orbit raising. The FCC authorisation also permits use of the beacon in the unlikely event of an emergency after a satellite has reached its authorised operational altitude, providing one possible context for some detections outside the orbit-raising phase. Nevertheless, the available records cannot establish whether these transmissions were associated with emergencies or with other operational activities.

This finding has potential implications for low-frequency radio astronomy. If VHF emissions can occur after the orbit-raising phase, their potential interference with SKA-Low may need to be considered over longer periods of satellite operation, rather than primarily during the months following launch. However, the available detections, including those reported here, date from 2025 or earlier, and it remains unknown whether subsequent changes to satellite systems or operating practices have altered this behaviour. Moreover, the existing observations cannot determine what fraction of satellites emit these signals, how frequently transmissions occur, or their overall impact on astronomical observations. Systematic monitoring across different satellite generations and orbital phases, combined with well-characterised detection sensitivity, will be needed to assess the extent and long-term implications of this potential source of radio-frequency interference. As a next step towards mitigation, the recovered chirp structure could be used to identify contaminated time--frequency cells and compare the fraction of data flagged by conventional energy-threshold detection and waveform-assisted detection at comparable levels of residual interference.

Serendipitously, Event~1 was discovered during an investigation of aircraft transits, when a large language model (LLM) assisting with the examination of observation data drew attention to an unexpected signal and suggested a possible Starlink VHF origin. This initial suggestion motivated the subsequent investigation, while the identification itself was established through independent signal processing and orbital analyses. This experience suggests that incorporating LLMs into the monitoring and characterisation of low-frequency radio environments could offer new opportunities for identifying and interpreting unfamiliar signals.

\section{Conclusions}
\label{section7}

Using raw voltage observations from the 21CMA E08 pod on 15 April 2025 (UTC), we detect and successfully decode two Starlink VHF TT\&C packets at 137.055~MHz. Both exhibit the characteristic LoRa waveform structure and yield valid PHY headers and payload CRCs. Each packet lasts approximately 1.027584~s, with the two events separated by 99.987~s. A systematic search of the five coordinated VHF channels across all observing windows identifies no additional packets satisfying the adopted detection and decoding criteria. 

By interpreting the decoded payload of Event~1 as an ECEF state vector and performing a blind comparison with 11\,305 active satellites, we identify the transmitter as STARLINK-32838 (NORAD ID 62827). This identification is further supported by independent SupGP data and the differential Doppler shift between the two events. Orbital analysis shows that the detections occurred during an orbit-raising phase, while the inferred transmission cadence and observing geometry provide a plausible explanation for the limited number of detections. For Event~1, we estimate a flux density of approximately 20.6~MJy, equivalent to 9.23~MJy at 1000~km and an EIRP of approximately 48.3~mW. Although subject to uncertainty in the sidelobe response, this estimate is consistent with previous EDA2 measurements at the order-of-magnitude level.

Our re-analysis of previously reported EDA2 detections further indicates that Starlink VHF emissions may not be confined to orbit raising, with detections also occurring during stable orbital phases and orbit lowering. Together, these results show that packet-level signal information, combined with orbital analysis, can improve the attribution of RFI from satellite emissions. For the specific 137.055~MHz Starlink VHF emission considered here, future monitoring should include satellites across different orbital phases rather than focus only on orbit raising. The prevalence, operational circumstances, and long-term impact of this emission remain to be established.


\section*{Data availability}

The two raw voltage files containing the detected events are available on Zenodo at \url{https://doi.org/10.5281/zenodo.23309576}. The analysis code and results are available at \url{https://github.com/JinYi1108/21CMA_Starlink_VHF}.

\begin{acknowledgements}
Quan Guo acknowledges the support of the NSFC under Grant No. 12673114. Qian Zheng acknowledges the support of the National Natural Science Foundation of China (NSFC) under Grant No. 12673113. Yun Yu acknowledges support from the Key Program of the National Natural Science Foundation of China (Grant No. 12433012).

\end{acknowledgements}

%

\bibliographystyle{aa}
\bibliography{references}

\end{document}